\documentclass[conference]{IEEEtran}
\IEEEoverridecommandlockouts
\usepackage{cite}
\usepackage{amsmath,amssymb,amsfonts}
\usepackage{algorithmic}
\usepackage{graphicx}
\usepackage{textcomp}
\usepackage{xcolor}
\def\BibTeX{{\rm B\kern-.05em{\sc i\kern-.025em b}\kern-.08em
    T\kern-.1667em\lower.7ex\hbox{E}\kern-.125emX}}
\usepackage{multirow}
\usepackage{multicol}
\usepackage[utf8]{inputenc}
\usepackage{enumitem}

\usepackage{xcolor}

\begin{document}

\title{Semantic Features Aided Multi-Scale Reconstruction of Inter-Modality Magnetic Resonance Images
}
\author{\IEEEauthorblockN{Preethi Srinivasan, Prabhjot Kaur}
\IEEEauthorblockA{\textit{School of Computing and Electrical Engineering} \\
\textit{Indian Institute of Technology, Mandi, India}\\
s18001,prabhjot\_kaur@students.iitmandi.ac.in}
\and
\and
\IEEEauthorblockN{Aditya Nigam, Arnav Bhavsar}
\IEEEauthorblockA{\textit{School of Computing and Electrical Engineering} \\
\textit{Indian Institute of Technology, Mandi, India}\\
aditya,arnav@iitmandi.ac.in}
}
\maketitle

\begin{abstract}
Long acquisition time (AQT) due to series acquisition of multi-modality MR images (especially T2 weighted images (T2WI) with longer AQT), though beneficial for disease diagnosis, is practically undesirable. We propose a novel deep network based solution to reconstruct T2W images from T1W images (T1WI) using an encoder-decoder architecture. The proposed learning is aided with semantic features by using multi-channel input with intensity values and gradient of image in two orthogonal directions. A reconstruction module (RM) augmenting the network along with a domain adaptation module (DAM) which is an encoder-decoder model built-in with sharp bottleneck module (SBM) is trained via modular training. The proposed network significantly reduces the total AQT with negligible qualitative artifacts and quantitative loss (reconstructs one volume in $\sim$1 second). The testing is done on publicly available dataset with real MR images, and the proposed network shows ($\sim$1dB) increase in PSNR over SOTA.
\end{abstract}

\begin{IEEEkeywords}
Inter-modality, Transformation-Learning, Encoder-decoder, T1-T2, Reconstruction
\end{IEEEkeywords}

\section{Introduction}
\label{sec:intro}
Distinct but related information from multi-modality MR images such as T1W, T2W, FLAIR, proton density weighted (PDw), functional-MRI, diffusion-MRI, etc., provides diagnosis benefits. However, the sequential acquisition of such images increases acquisition time which is generally undesirable. In this context, the T2W images require a much longer time to acquire as compared to T1W images which makes the acquisition procedure longer and thus prone to increased motion artifacts as well as cost. \color{black}Many acceleration methods for MR acquisition rely on undersampling the k-space, but suffer from trade-off between acquisition time and quality of MR images. For instance, acquisition of T1W and T2W images takes $\sim$10 minutes and may take $\sim$4-6 minutes for undersampled T2W k-space (1/8 samples) along with T1W, but the quality of such T2WI is too low to be used for diagnostic purposes. These issues demand post-capture algorithms which can efficiently and reliably map an image from one modality (e.g. T1W) with or without the the undersampled version of T2W to the other (T2W), obliviating the need to acquire the T2WI, \color{black} in lesser time than the scanner's acquisition time. Proposed work takes $\sim$1 second for the reconstruction of a complete T2W volume), \color{black} and thus reducing the overall acquisition time is the prime focus of this paper. 

\textbf{Related Work:} The existing approaches model such reconstruction problem as, learning a transformation between spaces spanned by acquired modality and target modality images in various frameworks based on sparse representation and neural network models~\cite{crossTMI,ding2018}. The synthesis of PDw and T2WI has been reported in~\cite{crossTMI} via learning the transformation along with dictionaries in the sparse representation framework. An unsupervised method to reconstruct the T2WI from T1WI is proposed in~\cite{unsupervisedInter} which works by selecting the best candidate from multiple target modality candidates via maximizing a global mutual information cost function. 
Of late, the accelerated reconstruction of MR images as well as the prediction of contrast for input images has been extensively explored using deep learning techniques~\cite{contrastmri,deepinMR}. 
Considering the importance of complementary information present in different modalities, few samples of k-space for T2WI, along with the complete T1WI, are for the first time, motivated to be utilized in construction of T2WI from given T1WI using DenseNet~\cite{ding2018}. 

\begin{figure*}[!ht]
\centering
\includegraphics[width=\linewidth,height=0.42\linewidth]{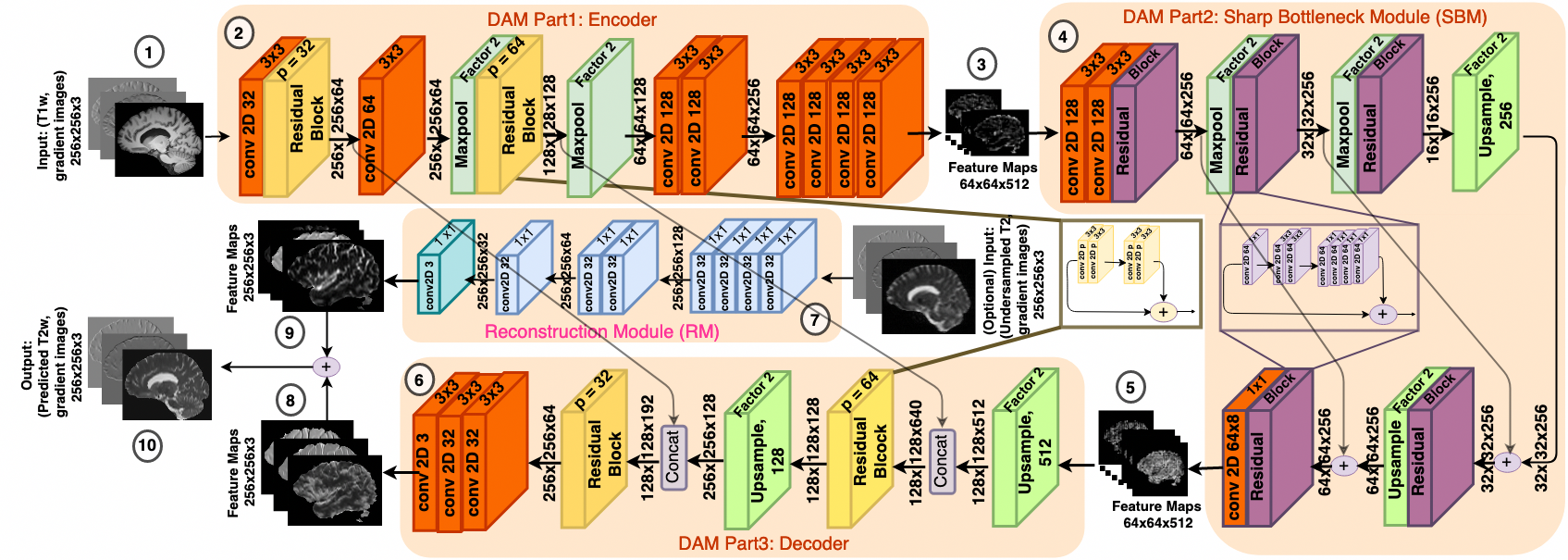}
\vspace{-0.8cm}
\caption{\small{Architectural details of proposed network. (Zoom for better visualization)}}
\vspace{-0.7cm}
\label{fig:network}
\end{figure*}

\textbf{Problem Statement: }Unlike many of the existing methods, the focus of this paper is to reconstruct T2WI from the given T1WI without any requirement for anatomical prior information. Since the quality of MR images is subjectively defined by clear tissue boundaries evaluated at various scales and a high SNR, we further suggest that it is important to process the brain MR image at multiple scales i.e. processing of micro as well as macro image details. Further, to provide a realistic solution for reconstruction of the T2WI, we focus on simpler network designs which impose less computational burden and yet is able to provide better reconstruction quality.

\color{black}
It has been suggested that the representative features for images can be learned using encoder-decoder networks, and is thus chosen as basis for the proposed work.
The overall encoder-decoder network along with stacked SBMs is named as the domain adaptation module (DAM) in this work. 

\color{black}The major contributions for proposed work are (i) Introducing RM network to incorporate the undersampled k-space information of T2WI to aid the DAM network for better learning. (ii) Image gradients in two orthogonal directions is fused in the input and ground truth for defining multi-channel loss, used for regularized network convergence. (iii)
The vanishing gradient problem is dealt locally in the network using residual networks and the stacked SBMs help in feature extraction at multiple scales. (iv) The proposed approach outperforms the existing algorithm~\cite{ding2018} in qualitative as well as quantitative measures.

\section{Proposed Work}
The proposed network (as shown in Fig.~\ref{fig:network}) consists of two major modules namely Domain adaptation module (DAM) and Reconstruction module (RM). The DAM reconstructs the T2WI from the corresponding T1WI (with two gradient images) using an encoder-decoder model. To generalize and enhance the learning ability, features encoded by encoder are further downsampled to very low dimensional space and immediately upsampled to learn the transformation at a finer scale. Such downsampling is done using stacked Sharp Bottleneck Modules (SBM), which is included as a part of the DAM. Under-sampled T2W k-space samples can also be utilized for enhanced learning and passed through convolutional layers of RM, till the output of DAM. 
Finally, the output of both RM and DAM modules is element wise added to produce the output of proposed network. The predicted T2W (along with two gradient images) and ground truth T2WI (with two gradient images) are compared using root mean square error (RMSE) loss for image similarity, and is computed only on the non-zero pixel values to emphasize the error only for brain region and to avoid 
vanishing gradients. Loss is back propagated using partial derivatives over addition into both the DAM and RM to improve learning.\color{black} We discuss some salient aspects of our approach 
with details and justification.

\textbf{Multi-Channel Input: }
Edges represent important semantic details in images, and it benefits to explicitly consider them while addressing the reconstruction problem for  
good localization of the structural details. We use a three channel input i.e. one channel contains the intensity values T1W image and other two channels contain the gradients of first channel in two orthogonal directions. The loss is computed over all three channels and thus the network estimates the intensity T2WI which has similar gradient profiles as of original T2WI (provided while training). In this way the network is regularized via gradients semantics that encodes edges and, thus forced to implicitly learn the semantically meaningful transformation which satisfies such mapping. 

\textbf{Reconstruction Module (RM) with DAM: } Although using only the DAM yields good quality reconstruction of T2WI, it does not use any prior information about the appearance of T2WI. To use such prior information (optional, if undersampled T2W information is available), we propose to augment DAM with a RM. RM can play a role of regularizing the optimization of DAM to learn a better transformation. 
The idea behind using the (undersampled) T2WI as a prior via RM instead of direct connection is that 
the convolutional layers of RM tend to learn the better reconstruction of T2WI at its own end (can be seen in activation maps in Fig.~\ref{fig:network}). As the output of RM and DAM modules are connected, so RM part reinforces the DAM to learn the transformation better, as the final estimate of the T2WI is now not just via the DAM, but is also constrained via the RM.

\textbf{Sharp Bottle Neck Module: }
\color{black}
The ability of a network to learn a better transformation between input and output image also depends upon the learning of the features at a variety of scales, so that the mapping is learnt between all such features. 
In MR images there exist many major and minor image details which play vital role in clinical diagnosis. We use max-pool functions in encoder-decoder network which enables the downsampling of images to provide features at various scale while preserving the semantic information and removing redundant values. SBM which downsamples and upsamples the features is arranged in cascaded fashion one after the other to re-evaluate the significance of the obtained global features after each SBM. \color{black}The proposed stacked SBM structure is inspired by the idea of repeated bottom-upsampling the images to obtain better features~\cite{hourglass}. It has been observed to introduce more non-linearity by churning out better multi-resolution features. 
However, the proposed work differs from ~\cite{hourglass} as we restrict the number of bottleneck modules to 2 instead of multiple as in~\cite{hourglass}, because MR images possess largely similar structure and thus superior features can be obtained using two bottleneck modules itself. 
Further, in order to estimate the T2W without any loss of image detail information in SBM (due to downsampling), skip connections are used to connect encoding and decoding part of SBM and residual connections are utilized in encoding and decoding part of the DAM. They help in retaining information at the global and local features respectively, thus allowing easy flow of gradients.

\begin{figure*}[!htp]
    \centering
    \includegraphics[height=0.32\linewidth,width=\linewidth]{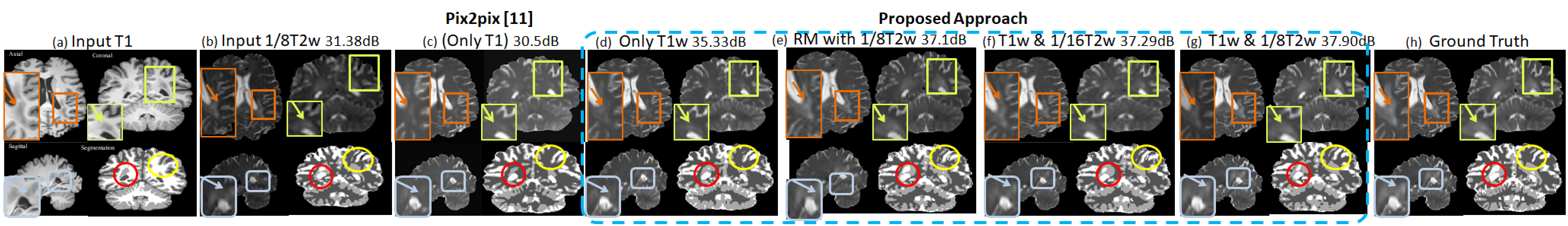}
    \vspace{-0.9cm}
    \caption{\small{Reconstructed images from proposed network and pix2pix~\cite{pix2pix} network with different kinds of inputs. Each block shows views of output image in (clockwise) Axial plane, Coronal plane, Segmentation map of Coronal plane, Sagittal plane.}}    \label{fig:qualitydataset1}
    \vspace{-0.55cm}
\end{figure*}

\textbf{Modular Training: }
The network is adapted for MRI via effective modular training. A sequence of steps as given below, makes the training of network efficient as well effective: 
(i) Firstly, only the encoder and decoder models of DAM are trained for reconstructing T2WI from only T1WI. \color{black} For the case where undersampled T2WI are to be used in RM, the encoder, decoder as well as RM is trained in first step. \color{black} (ii) Secondly, one SBM has been added between the trained encoder and decoder. The weights of trained encoder and decoder (RM also in alternative case) are loaded and frozen. Then, only the SBM is trained. Although the Loss back propogates through all the 3 parts (Encoder-SBM-Decoder), only the weights of the SBM will be updated according to the gradients wrt the Loss (iii) Finally, copy of first SBM is made and two of them are stacked between encoder and decoder to form the full DAM. Now, the weights of Encoder-Decoder from Step 1 are loaded and weights of SBM from step 2 are loaded on itself and its copy. Loss is back-propagated and weights are updated accordingly in all the 4 pieces (Encoder-SBM-SBM-Decoder) and in all the 5 pieces in case of DAM with RM. Training is done in an end to end fashion in this step. Benefits of modular training are:
    (a) 
    Learning a complex transformation involving large number of parameters ($\sim$6.9M) is difficult. 
    So the parameters are learned in parts at first to achieve reasonably good quality piece wise solution, and then fused together to improve the solution further. 
    (b) It also helps the learned parameters to be more robust to the variations (contrast, scales of image details etc) in input scenario.

\section{Experimental Results}
To demonstrate the performance of the proposed work, experiments are performed on publicly available dataset with real MR images, having T1W-T2W paired images for 5 subjects with spatial resolution $336\times336\times261$~\cite{commowick2016msseg}. Leave-one-out cross-validation technique is used for evaluation 
to compare the performance with existing methods.

\color{black}
The undersampling procedure in this work is adopted from existing works~\cite{ding2018}, for the fair comparison. The k-space samples are first undersampled retrospectively followed by choosing a fraction of samples from central low frequency components. For example, 1/8 T2W means the T2WI generated from k-space with $\frac{1}{8}^{th}$ low frequencies. Such a  strategy is realistic as only the central part of k-space has to be sampled 
and thus quality of undersampled image deteriorates. The 1/8 T2W used in proposed work has 31.38dB PSNR which is lower than 1/8 T2W used in~\cite{ding2018} with 32.4dB, and thus can be used for fair comparison. 


\color{black}
\textbf{Qualitative Comparison of Reconstructed Images: }
The T2WI are reconstructed using (i) only T1WI, (ii) Only 1/8 k-space sampled T2WI, (iii) T1WI along with T2WI obtained from 1/8 k-space samples, and (iv) T1WI with 1/16 T2WI. The reconstruction results of proposed work are compared with style transfer algorithm for images called pix2pix~\cite{pix2pix} and with~\cite{ding2018}. Though the proposed work addresses 2D image processing (in axial plane) which inherently assumes the independence among slices (which is not true at least for neighbouring slices), but we here show the reconstruction in all three planes for a randomly selected image in Fig.~\ref{fig:qualitydataset1} to better evaluate the reconstruction quality visually. In addition to the quality comparison in terms of better estimation of image details and reduced artifacts while reconstruction, we also focus on the utility of reconstructed images for post-processing applications and thus segmentation maps for cerebral spinal fluid (CSF), white matter (WM) and gray matter (GM) are shown using three distinct gray levels in Fig.~\ref{fig:qualitydataset1}. The image details with abnormality is shown in zoomed window. PSNR values of each of the slices are also mentioned above each block. It can be observed that the image reconstructed by pix2pix \cite{pix2pix}, 
does not reconstruct the abnormal image detail well (in all three planes) but the proposed approach with T1WI input is able to reconstruct the details better in axial plane. Also, the pix2pix in unable to provide accurate segmentation labels and in fact tends to produce artifact by removing the gray segmentation label, can also be seen in red and yellow circles, as compared to proposed method. Such image details comprise complementary information from other modality and thus RM is able to better reconstruct such image details.  
The zoomed detail in axial slice shows crucial information which is difficult to perceive in both input T1WI and corresponding reconstructed image, but is reconstructed well when degraded version of T2WI is given in addition through RM.

\begin{table}[]
\centering
\caption{\small{Quantitative comparison with existing work~\cite{ding2018}.}}

\resizebox{8.5cm}{!}{
\begin{tabular}{|p{1cm}|c||p{1.9cm}|p{1.9cm}|p{1.9cm}|}

\hline
Metric                     & Method   & Reconstructed T2 with only T1 & Reconstructed T2 with 1/8 T2 and T1 & Reconstructed T2 with 1/16 T2 and T1 \\ \hline \hline
\multirow{2}{*}{PSNR} & DenseNet~\cite{ding2018} & 30.60                         & 36.90                               & 34.3                                 \\ \cline{2-5} 
                           & Proposed & \textbf{34.07}                         & \textbf{37.30}                               & \textbf{36.50}                                \\ \hline
\multirow{2}{*}{MAE}       & DenseNet~\cite{ding2018} & 33$\times$10$^{-3}$           & 14$\times$10$^{-3}$                 & 19$\times$10$^{-3}$                  \\ \cline{2-5} 
                           & Proposed & \textbf{5.01$\times$10$^{\textbf{-3}}$}           & \textbf{3.46$\times$10$^{\textbf{-4}}$}                 & \textbf{3.87$\times$10$^{\textbf{-4}}$}                  \\ \hline \hline
\multirow{2}{*}{\textbf{Data-II}} & PSNR & 32.13                         & 33.08                               & 32.96                                 \\ \cline{2-5} 
                           & MAE & {7.81$\times$10$^{\textbf{-3}}$}           & {6.86$\times$10$^{\textbf{-3}}$}                 & {7.13$\times$10$^{\textbf{-3}}$}                            \\ \hline
\end{tabular}}
\vspace{-0.6cm}
\label{tab:quanti}
\end{table}
\textbf{Quantitative Comparison of Reconstructed Images: }
Since only~\cite{ding2018} has attempted to use the partial information of T2 k-space and is proved to perform better than U-Net, we only compare our results with~\cite{ding2018}. The quantitative comparison is tabulated in Table~\ref{tab:quanti}. It has to be noted that the dataset and pre-processing steps are kept minimal as well as same in proposed work to that in ~\cite{ding2018}, so the PSNR values mentioned in manuscript are quoted for~\cite{ding2018}. \color{black}It can be observed that even the 1/8 T2W used in proposed network has lesser PSNR as compared to 1/8 T2W in~\cite{ding2018}, the proposed work outputs T2WI which has higher PSNR as compared to~\cite{ding2018}, signifying better transformation learning by proposed network. We also provide comparisons between our approach and pix2pix~\cite{pix2pix} for only T1W as input because there was option provide undersampled image as guidance. The best PSNR obtained using \textit{\textbf{{pix2pix~\cite{pix2pix} is 30.1dB}}} which is lesser than PSNR of the \textit{\textbf{proposed work which is 34.07dB}}.

\textbf{Performance on other dataset: }To prove the feature learning ability by the proposed network on other datasets, 3T MRI images from HCP (45/6/6 subjects for training/validation/test) were used (dataset-II)~\cite{dataset1}. The PSNR values obtained for reconstructed T2WI are quoted in Table.~\ref{tab:quanti}, 
which indicate the similar quality obtained for these datasets too. 

\begin{figure}
    \centering
    \includegraphics[width=\linewidth,height=0.34\linewidth, clip=true,trim={0cm 0cm 2.25cm 0.56cm}]{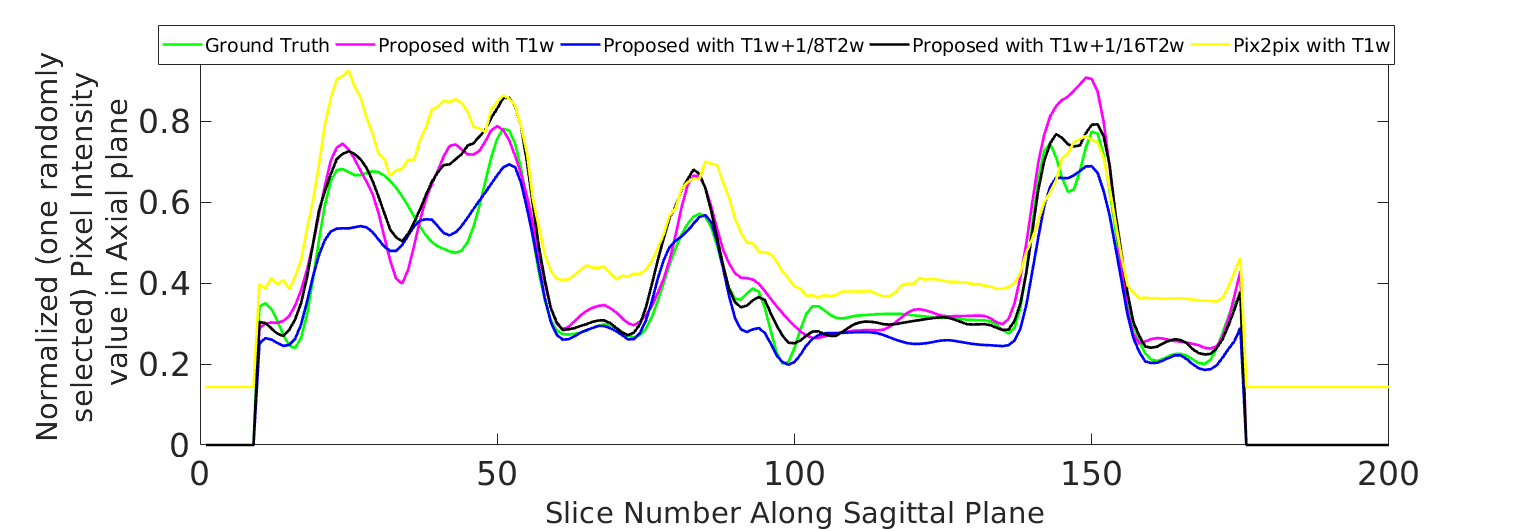}
    \vspace{-0.75cm}
    \caption{\small{Slice level inconsistency while 2D reconstruction}}
    \label{fig:lineplot}
    \vspace{-0.5cm}
\end{figure}

\textbf{Slice-wise Inconsistency: }Further to check for inconsistency across the slices while reconstructing 2D slice in 3D volume data, we randomly choose a pixel in axial slice and compared its variation in sagittal plane with original T2WI, shown in Fig.~\ref{fig:lineplot}. It can be seen that the proposed approach is able to follow the smooth changes such as from 100 to 150 slices, however, lacks in case of abrupt changes such as 0 to 50 slices. The pix2pix approach provides relatively non-smooth variation overall and does not reconstruct the sudden change such as in the peak of 150$^{th}$ slice.



\textbf{Ablation Study: }
The emphasis of each component of proposed work i.e., encoder-decoder network, SBM, RM is evaluated by reconstructing images from respective modules. The obtained PSNR values for the reconstruction using different components are tabulated in Table~\ref{tab:ablation}. It can be seen that the PSNR obtained by DAM network without SBMs is less than DAM network with SBMs, indicating effective learning by SBMs. Further, it has been observed in experiments that the SBMs helps in smoother convergence, which is otherwise shaky while learning with only encoder-decoder and RM. 



\section{Summary}
This network stems from encoder-decoder architecture, embeds SBM in between encoder and decoder for scale variability in feature extraction as well as for guided convergence, and RM is connected in parallel which leads to a regularized optimization. The proposed network yields relatively better reconstruction results, and can be easily utilized for FLAIR or any similar reconstruction.

\begin{table}[]
    \centering
    \caption{Quantitative Analysis (Ablations)}
\resizebox{8cm}{!}{
    \begin{tabular}{|c|c|c|}
    \hline
        \textbf{Training Module}. & \textbf{T1W input} & \textbf{T1W+1/8T2W input }\\ \hline
        DAM & 32.08 & 33.10\\ \hline
        DAM+1HG & 31.87 & 33.13\\ \hline
        DAM+2HG & 32.16 & 33.08\\ \hline
    \end{tabular}}
    \label{tab:ablation}
    \vspace{-0.2cm}
\end{table}

\bibliographystyle{IEEEbib}
\bibliography{refs.bib}

\end{document}